\documentclass[preprint,showpacs,preprintnumbers,amsmath,amssymb]{revtex4}

\usepackage{graphicx}
\usepackage{dcolumn}
\usepackage{bm}

\newcommand{\ve}[1]{\mbox{\boldmath$ #1 $}}

\begin{document}


\title{Radiative Effects on Particle Acceleration in Electromagnetic Dominated Outflows}

\author{Koichi Noguchi}
%
\affiliation{
Rice University,
Houston, TX 77005-1892
}%
\author{Kazumi Nishimura}%
\affiliation{%
AdvanceSoft, Tokyo Japan
}%

\author{Edison Liang}
\affiliation{
Rice University,
Houston, TX 77005-1892
}%

\date{\today}

\begin{abstract}
Plasma outflows from gamma-ray bursts (GRB), pulsar winds, relativistic
jets, and ultra-intense laser targets radiate high energy photons. However, radiation damping
is ignored in conventional PIC simulations.
In this letter, we study the radiation damping effect on particle acceleration via Poynting fluxes
in two-and-half-dimensional particle-in-cell (PIC) plasma simulation of electron-positron plasmas.
Radiation damping force is self-consistently calculated for each particle and  
reduces the net acceleration force.
The emitted radiation is peaked within a few degrees from 
the direction of Poynting flux and strongly linear-polarized.
\end{abstract}

\maketitle

When charged particles suffer extreme acceleration, radiation loss and damping become 
important in the plasma energetics and dynamics. This is especially true in the ultra-relativistic 
regime, where radiation damping can severely limit individual particle acceleration.  
However, conventional Particle-in-Cell (PIC)\cite{bird85,lang76,lang92,nish02} simulations of collisionless plasmas have not 
included radiation effects. 

One of such an example is high-energy astrophysical phenomena, such as pulsars, blazars and gamma-ray burst(GRB) emissions.
There are two competing paradigms for the origin of the prompt GRB emissions: hydrodynamic internal shocks\cite{pira00,paes04,sil03,nis03,mesz02} versus Poynting 
fluxes\cite{lyut01}.  Both pictures require the rapid and efficient 
acceleration of nonthermal electrons to high Lorentz factors in moderate magnetic fields to 
radiate gamma-rays. In the Poynting flux scenario, long-wavelength electromagnetic (EM) 
energy can be directly converted into gamma-rays using the electrons or electron-positron pairs as 
radiating agents. Such Poynting flux may originate as hoop-stress-supported magnetic jets driven by strongly magnetized accretion onto a nascent blackhole, or as transient millisecond magnetar wind, 
in a collapsar event\cite{zhan03} or in the merger of two strongly magnetized compact objects\cite{ruff03}. 

The recent discovery of the diamagnetic relativistic pulse 
accelerator (DRPA)\cite{lian03,lian04,lian05,nish03,nish04}, 
in which intense EM pulses imbedded inside an overdense 
plasma (EM wavelength $\lambda \gg$ plasma skin depth $c/\omega_{pe}$) capture and accelerate 
surface particles 
via sustained in-phase Lorentz forces when the EM pulses try to escape from the plasma, is 
particularly relevant to the Poynting flux scenario of GRBs. 
Liang \& Nishimura \cite{lian04} recently showed that DRPA reproduces from first-principles many of the unique features 
of GRB pulse profiles, spectra and spectral evolution. DRPA-like Poynting flux acceleration may occur when a Poynting jet head emerges from the surface of a collapsar, or when a new born magnetar wind blows out the progenitor envelope.

In this article we report PIC simulation results of particle acceleration driven 
by EM-dominated outflow (Poynting flux), using a newly developed 2-1/2-D code that includes 
self-consistent radiation damping.  
We compute the radiation-damped plasma and field evolutions, as well as observable radiation output, which may be applicable to both laser experiments and astronomical phenomena. For example, focusing ultra-intense short pulse lasers\cite{mour98} into small focal spots, the intensity may exceed $10^{22}$ W/cm$^2$\cite{aoya03,esir04}, and to the Schwinger limit, $10^{29}$ W/cm$^2$ with couner-propagating pulses\cite{bula03}. Under such conditions, electrons can be accelerated to relativistic speed, and the effect of the radiation damping decreases the efficiency of the laser heating\cite{zhid02}. In future, we will apply the code to such a case.


High frequency radiation with $\lambda \ll \lambda_D \equiv c/\omega_{pe}$ cannot be included as 
part of the self-consistent EM fields in PIC simulations since the Maxwell-solver~\cite{bird85} 
cannot capture gradients $>\lambda_D^{-1}\equiv \omega_{pe}/c$, where $\omega_{pe}=\sqrt{4\pi n e^2/m_e}$ is the electron plasma frequency.

Relativistic particles can emit up to the critical frequency 
$\omega_c=3\gamma^2\Omega_{ce}$,
where $\gamma=E/m_e c^2=1/\sqrt{1-v^2/c^2}$ is the Lorentz factor 
and $\Omega_{ce}=eB/(m_e c)$ is the electron gyro-frequency.
The ratio of the critical radiation wavelength $\lambda_c$ to $\lambda_D$ is given by
$\lambda_c/\lambda_D=(2\pi\omega_{pe})/(3\gamma^3\Omega_{ce})$, which is $\ll 1$
because $\omega_{pe}/\Omega_{ce}<0.1$ in EM-dominated cases and $\gamma\gg 1$.

To account for high-frequency radiation, we introduce a radiation damping
force in the form of the Dirac-Lorentz equation~\cite{land75,rohr01,koga04}.
The classical damping force term $\ve{f}_{rad}$ is given by (see Ref.~4 for detailed calculations)
\begin{eqnarray}
\ve{f}_{rad}&=&\frac{2e}{3\Omega_{ce}}k_{rad}\times\nonumber\\
&&\!\!\!\!\!\!\!\!\left\{\gamma\left[\left(\frac{\partial}{\partial t}+\ve{v}\cdot\nabla\right)\!\!\ve{E}
+\frac{\ve{v}}{c}\times\!\!\left(\frac{\partial}{\partial t}+\ve{v}\cdot\nabla\!\!\right)\!\!\ve{B}\right]\right.\nonumber\\
&&\!\!\!\!\!\!\!\!+\frac{e}{mc}\left[\ve{E}\!\times\!\ve{B}+\frac{1}{c}\ve{B}\!\times\!(\ve{B}\!\times\!\ve{v})
+\frac{1}{c}\ve{E}(\ve{v}\cdot\ve{E})\right]\nonumber\\
&&\!\!\!\!\!\!\!\!\left.-\frac{e\gamma^2}{mc^2}\ve{v}
\left[\left(\ve{E}+\frac{1}{c}\ve{v}\times\ve{B}\right)^{\!\!2}\!\!-\frac{1}{c^2}(\ve{E}\cdot\ve{v})^2\right]\right\},\hspace{3mm}
\label{radf2}
\end{eqnarray}
where $\ve{v}$ is the velocity, and $\ve{E}$ and $\ve{B}$ are the self-consistent 
electric and magnetic fields. Here we introduce a non-dimensional 
factor $k_{rad}$ given by
\begin{equation}
k_{rad}=\frac{r_e\Omega_{ce}}{c}=1.64\times10^{-16}\times B\mbox{(gauss)},
\end{equation}
where $r_e=e^2/(mc^2)$ is the classical electron radius. 

The first square bracket term of Eq.~(\ref{radf2}) represents the radiation 
damping due to the ponderomotive force acceleration. 
The third square bracket term is Compton scattering by large scale $(\lambda >\lambda_D)$ 
electromagnetic field
which reduces to Thomson scattering in the classical limit~\cite{rybi79}. 
We note here that Compton scattering with high frequency radiation ($\lambda\ll\lambda_D$) is not 
included in Eq. (\ref{radf2}).

We restrict ourselves to stay below the quantum-limit, 
$\hbar\Omega_{ce}\leq m_e c^2$ or $B\leq 4.4\times 10^{13}$ gauss, which corresponds 
to $k_{rad}=\leq 7.2\times10^{-3}$, otherwise formula (\ref{radf2}) fails. This is the case
for magnetars ($k_{rad}\simeq 10^{-2}$).
We choose $k_{rad}$ from zero to $10^{-3}$ in the simulation to maximize the radiation 
effect and run our simulations until $\tau_{sim}\geq 10^4\Omega_{ce}^{-1}$ so that $|\ve{f}_{rad}|\tau_{sim}\simeq|\ve{F}_{ext}|\Omega_{ce}^{-1}$.

We use the explicit leap-frog method 
for time advancing \cite{bird85}. 
Spacial grids for the fields are uniform in both $x$ and $z$ directions, 
$\Delta x=\Delta z=\lambda_D$. The simulation domain in the $x\!-\!z$ plane is
$-L_{x}/2\leq x\leq L_{x}/2$ and
$0\leq z\leq L_{z}$ with a doubly periodic boundary condition in both directions.

As an example to highlight the effect of radiation by intense EM pulse, we here repeat the non-radiative simulations by Liang et.~al.~\cite{lian03} using the similar initial condition in the following simulations. The initial plasma is uniformly distributed 
at the center of the simulation box, $-6\Delta x<x<6\Delta x$ and $0<z<L_z$. 
The background uniform magnetic 
field $\ve{B}_0=(0,B_0,0)$ is applied only inside the plasma, so that the magnetic field freely 
expands toward the vacuum regions. 
We choose $L_x$ to be long enough so that plasma and EM wave never hit the boundaries 
in the $x$ direction within the simulation time. The initial plasma is assumed to be a spatially uniform relativistic Maxwellian, 
$k_BT_e=k_BT_p=1$MeV, where the subscripts $e$ and $p$ refer to electrons and positrons. 

The radiation damping force (\ref{radf2}) is calculated self-consistently and fully-explicitly as follows.
The velocity of each particle is updated each single time step $\Delta t$
 from $\ve{v}(t-\Delta t/2)$ to $\ve{v}(t+\Delta t/2)$, using the electromagnetic field
and $\ve{f}_{rad}$ at time $t$. All the terms in the equation (\ref{radf2}) can be calculated by
the ordinary leap-frog field solver\cite{bird85,lang76,lang92} except the displacement current $d\ve{E}/dt$, which is not given at time $t$. 
To calculate $d\ve{E}/dt$, particle position $x$ need to be updated from $t-\Delta t/2$ to $t$ using 
velocity $\ve{v}(t-\Delta t/2)$. Next, temporal current $\ve{J}_t(t)$ is calculated from 
$\ve{v}(t-\Delta t/2)$ and $\ve{x}(t)$. $\ve{J}_t$ has the first order accuracy in time.
Finally, the displacement current is calculated using the Maxwell equation, and the radiation damping 
force is calculated for each particle via Eq. (\ref{radf2}).
To update the velocity, we sum $\ve{f}_{rad}$ and the external force 
as a net 'acceleration' force, and apply the Boris rotation\cite{bird85}.

We run simulations for six different sets of parameters shown in Table~\ref{table1};  
$k_{rad}=0$, $10^{-4}$, $10^{-3}$ and $\omega_{pe}/\Omega_{ce}=0.1$, $0.01$.
We call Run A-C the weak magnetic field case and Run D-F 
the strong magnetic field case.

\begin{figure}
\includegraphics[width=\linewidth]{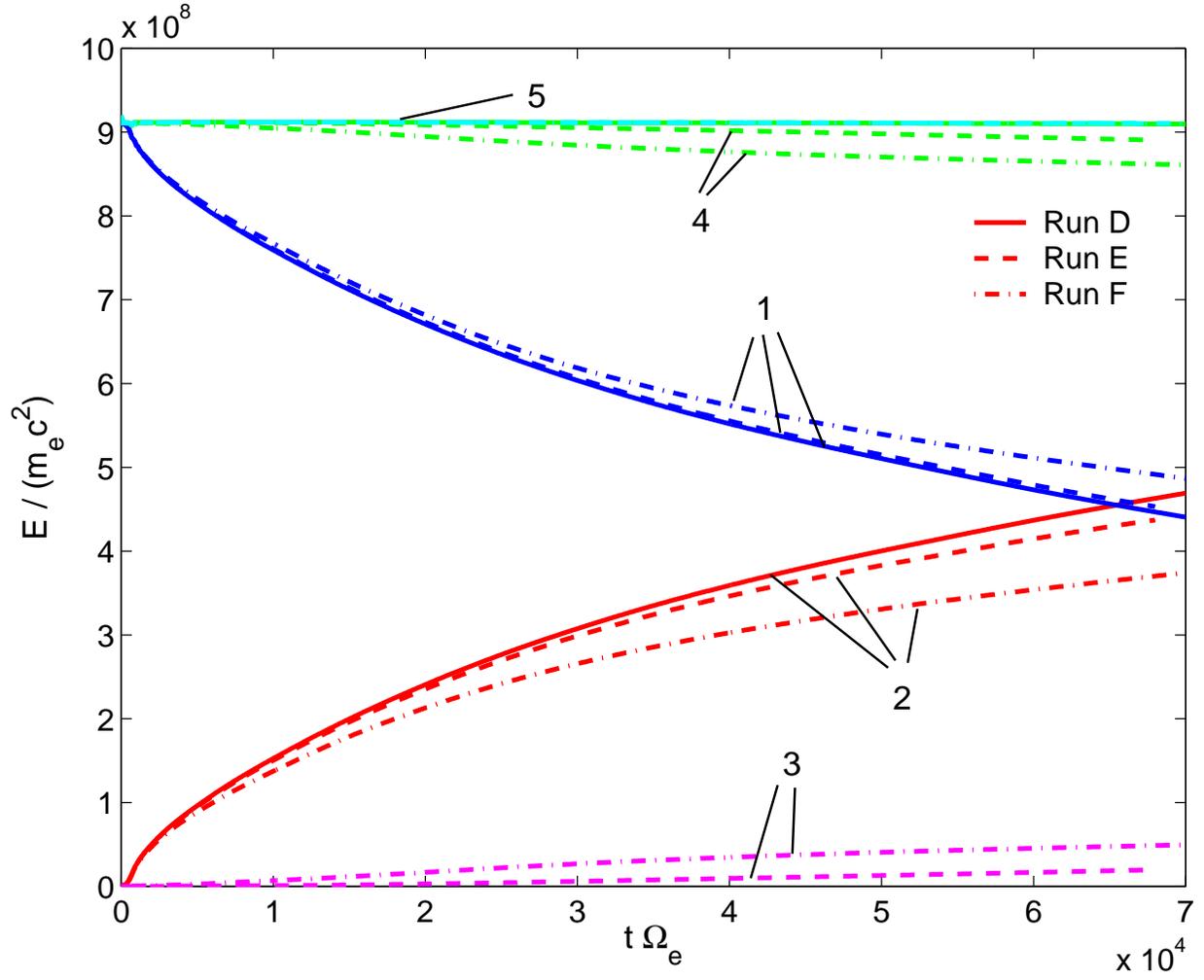}
\caption{\label{fig:eng} System-integrated energy in the electromagnetic field (1), 
particles (2), radiation loss (3), sum of field and particle energy (4), and total
energy (5) as functions of time for Run D (solid), E (dashed) and F (dash-dot).}
\end{figure}

First, we check total energy conservation for Run D,E  and F in Fig.~\ref{fig:eng}. 
In the radiative (RD) cases, 
the energy loss by the radiation $E_{rad}$  is obtained from the time integral
\begin{equation}
E_{rad}=\int_0^t\left(\sum_{e,p}\ve{v}\cdot\ve{f}_{rad}\right) dt.
\end{equation}
In the RD case, sum of $E_{kin}$ and $E_{fie}$ does not 
conserve (Line 4), but sum of $E_{kin}$, $E_{fie}$ and the radiation energy $E_{rad}$ (Line 3) conserves (Line 5),
indicating that the radiation damping force is self-consistently calculated. 
In all the RD cases, the energy is transferred from field to particle, and then radiation. 
Energy transfer, however, from field to particles becomes less efficient 
as radiation damping increases. Radiation damping suppresses the build-up of the
high energy tail of the particle distribution.

\begin{figure}
\includegraphics[width=\linewidth]{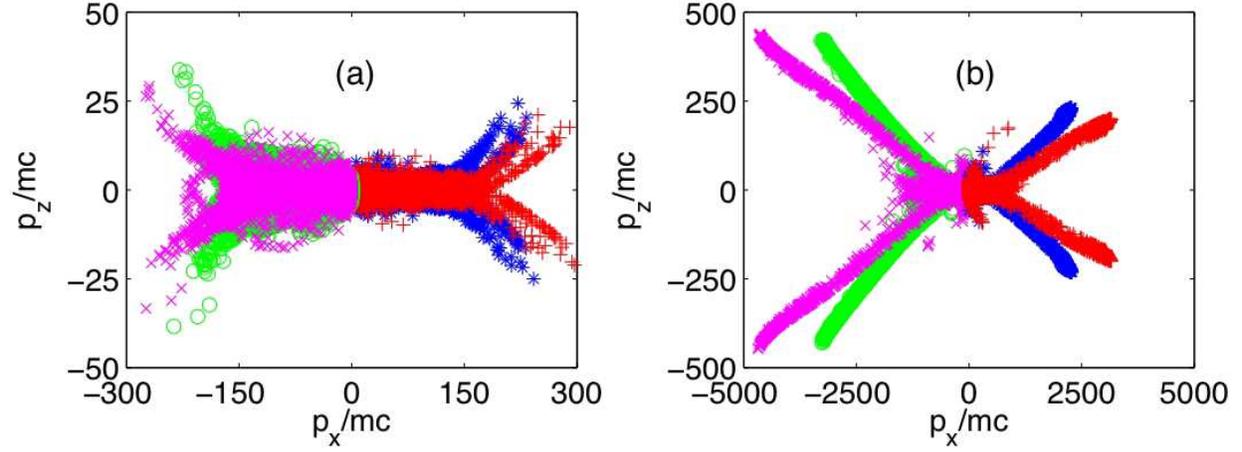}
\caption{\label{fig:phs}Comparison of momentum distribution of particles between Run A [(a), $P_x<0$] and Run C [(a), $P_x>0$]
at $t\Omega_{ce}=5000$ (circle, asterisk) and $t\Omega_{ce}=10000$ (x, plus), and
between Run .D [(b), $P_x<0$] and Run F [(b), $P_x>0$] at $t\Omega_{ce}=35000$ (circle, asterisk) and $t\Omega_{ce}=70000$
(x, plus). Results for $P_x>0$ and $P_x<0$ are identical in all cases, so only half momentum spaces are plotted. This shows that radiation damps both $P_x$ and $P_z$ but the ratio between them relatively unaffected, and that the damping increases with magnetic field.}  
\end{figure}

Figure \ref{fig:phs} shows the momentum distribution of particles for (a) Run A($P_x<0$) and Run C($P_x>0$), 
and (b) Run D($P_x<0$) and Run F($P_x>0$). 
The $\ve{v}\times\ve{B}$ force accelerates electrons and positrons in the same direction along the $x$ axis, 
whereas electric field accelerates electrons and positrons oppositely
along the $z$ direction, forming X shape distribution in the $p_x-p_z$ plane as a result.
The ponderomotive force $\ve{J}\times\ve{B}$ 
creates successive 'potential wells', which captures and accelerates co-moving 
particles in the $x$ direction. We emphasize here that there is no charge separation in the $x$ direction. Particle momenta in both $x$ and $z$ directions are radiated away 
in both weak and strong magnetic field RD cases. 
Radiation damping is more severe for higher $\ve{B}$. We also find that radiation enhances the bifurcation instability discussed by Liang and Nishimura\cite{nish04}.


Next, we compare the radiation power of RD cases with NRD cases.
For the NRD case, we estimate the radiation power using the relativistic dipole formula
\cite{rybi79}
\begin{equation}
\langle P\rangle=\frac{2e^2}{3m^2c^3}(F_\parallel^2+\gamma^2 F_\perp^2),\label{dipole}
\end{equation}
where $F_\parallel$ and $F_\perp$ are the parallel and perpendicular components of the force with
respect to the particle's velocity. 
The bracket $\langle\rangle$ indicates that we take the average 
of all the particles located within $30\lambda_D$ from the pulse front.  
We plot Runs A and D with $k=10^{-3}$ and $10^{-4}$ to compare with the RD cases.

For the RD ($k>0$) cases, the radiation power is calculated using the formula
\begin{equation}
\langle P\rangle=|\ve{f}_{rad}\cdot\ve{v}|.\label{landau}
\end{equation}
We compared the result of these two formulae for the RD cases, and they matched within the linewidth. 
Therefore, We only show the result of Eq. (\ref{landau}) here. 

\begin{figure}
\includegraphics[width=\linewidth]{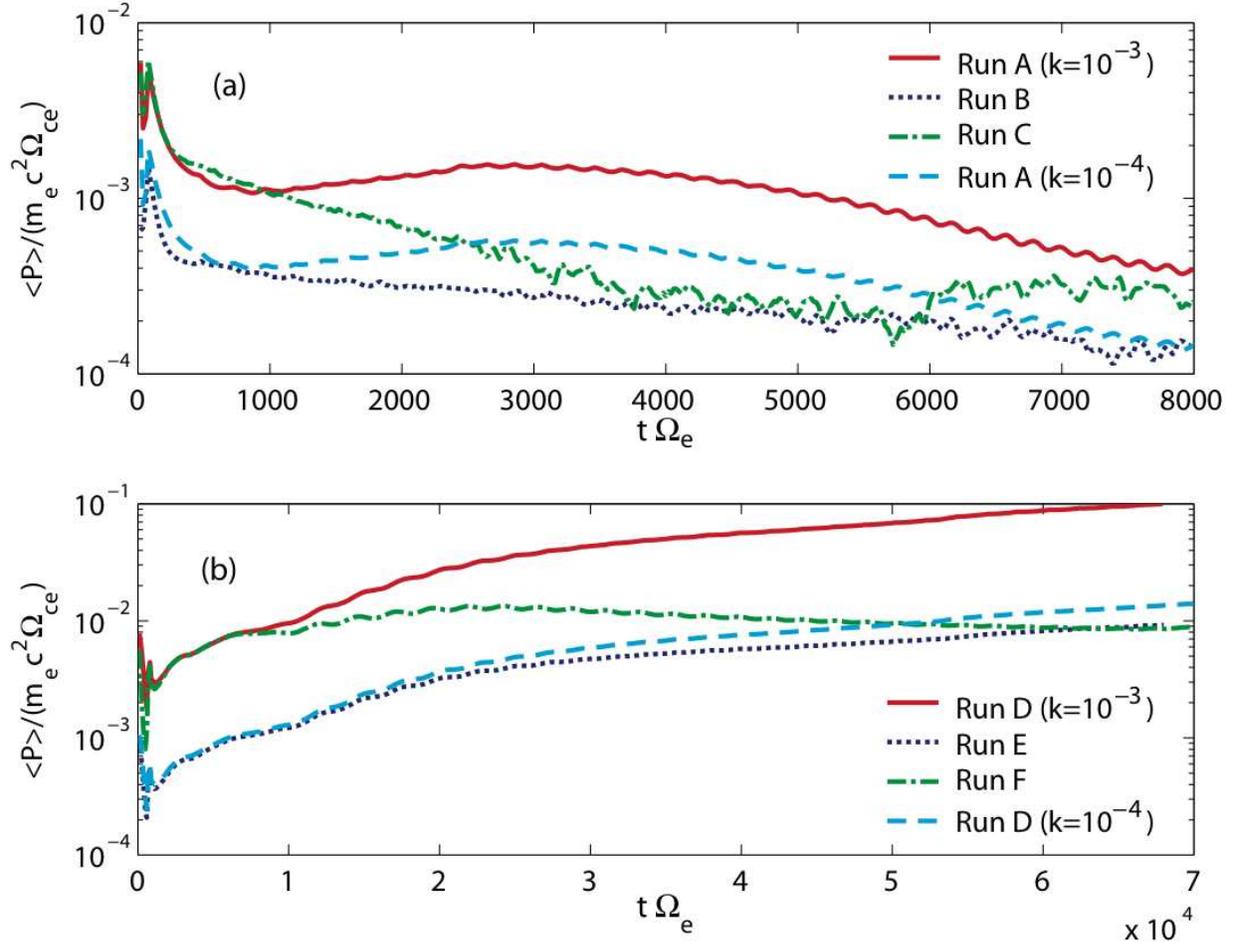}
\caption{\label{fig:pow} Average instantaneous radiation power from particles within $30\lambda_D$ 
from pulse front for (a) $\omega_{pe}/\Omega_{ce}=0.1$ (Runs B, C)  and (b) 0.01 (Runs E, F). 
Estimated power for NR cases (Runs A, D) are also shown for comparison. Radiation monotonically decreases in time, and reaches more or less constant in (b), whereas it still decreases in (a).}
\end{figure}

Figure \ref{fig:pow} shows the instantaneous radiation power $\langle P\rangle$ for all runs.
The initial peak around $t\Omega_{ce}=100$ is due to synchro-cyclotron emission \cite{rybi79},
and the estimated radiation power for the NRD case quantitatively matches with the RD case.
At later times ($t \Omega_{ce}>1000$), however, less power is radiated 
in the RD cases than NRD cases, because energetic particles are self-consistently
decelerated by the radiation damping in the RD cases.


Next, we calculate the self-consistent radiation field and 
its angular dependence directly.
Intensity $I$ and polarization $\Pi$ of the radiation received by the observer located at $\ve{x}$ 
are given by \cite{rybi79,jack75}
\begin{equation}
I(\hat{\ve{n}},\tau)
=\sum_i\left[
    \left|\ve{E}_i\right|^2
\right]_{\mbox{ret}},\label{inten}
\end{equation}
and
\begin{eqnarray}
E_y^2(\hat{\ve{n}},\tau)&=&\sum_{i,\tau}\left|
    \ve{E}_i\cdot\hat{\ve{y}}
\right|^2\!\!,\quad
E_z^2(\hat{\ve{n}},\tau)=\sum_{i,\tau}\left|
    \ve{E}_i\cdot\hat{\ve{z}}
\right|^2\!\!,\nonumber\\
U(\hat{\ve{n}},\tau)&=&2\sum_{i,\tau}
    (\ve{E}_i\cdot\hat{\ve{y}})(\ve{E}_i\cdot\hat{\ve{z}}),\label{EyEz}\\
\Pi(\hat{\ve{n}},\tau)&=&\frac{\sqrt{(E_z^2)^2+(E_y^2)^2-2E_y^2E_z^2+U^2}}
    {E_z^2+E_y^2},
\end{eqnarray}
where 
\begin{equation}
\ve{E}_i=\frac{e}{c}\frac{\hat{\ve{n}}\times[(\hat{\ve{n}}
    -\ve{\beta}_i)\times\dot{\ve{\beta}}_i]}
    {(1-\hat{\ve{n}}\cdot\ve{\beta})^3 R},
\end{equation}
is the radiated electric field from particle $i$ located at $\ve{r}$, 
$\hat{\ve{n}}$ is a unit vector in the direction of $\ve{x}-\ve{r}(\tau)$, 
$\ve{\beta}=\ve{v}(\tau)/c$, and $\dot{\ve{\beta}}=d\ve{\beta}/dt$.
We assume that $|\ve{x}|\gg|\ve{r}|$ so that $\hat{\ve{n}}$ is parallel to $\ve{x}$.
The summation in Eqs. (\ref{inten}) and (\ref{EyEz}) is 
evaluated at the retarded time $\tau=t-R/c$, where $R=|\ve{x}-\ve{r}|$. 
We take $\tau=0$ when the pulse front reaches the observer.
To specify the direction of the observer with respect to the $x$ axis,
we introduce $\theta$ and $\phi$ as
\begin{equation}
\hat{\ve{n}}=(\cos \theta \cos \phi, \cos \theta \sin \phi, \sin \theta).
\end{equation}

\begin{figure}
\includegraphics[width=\linewidth]{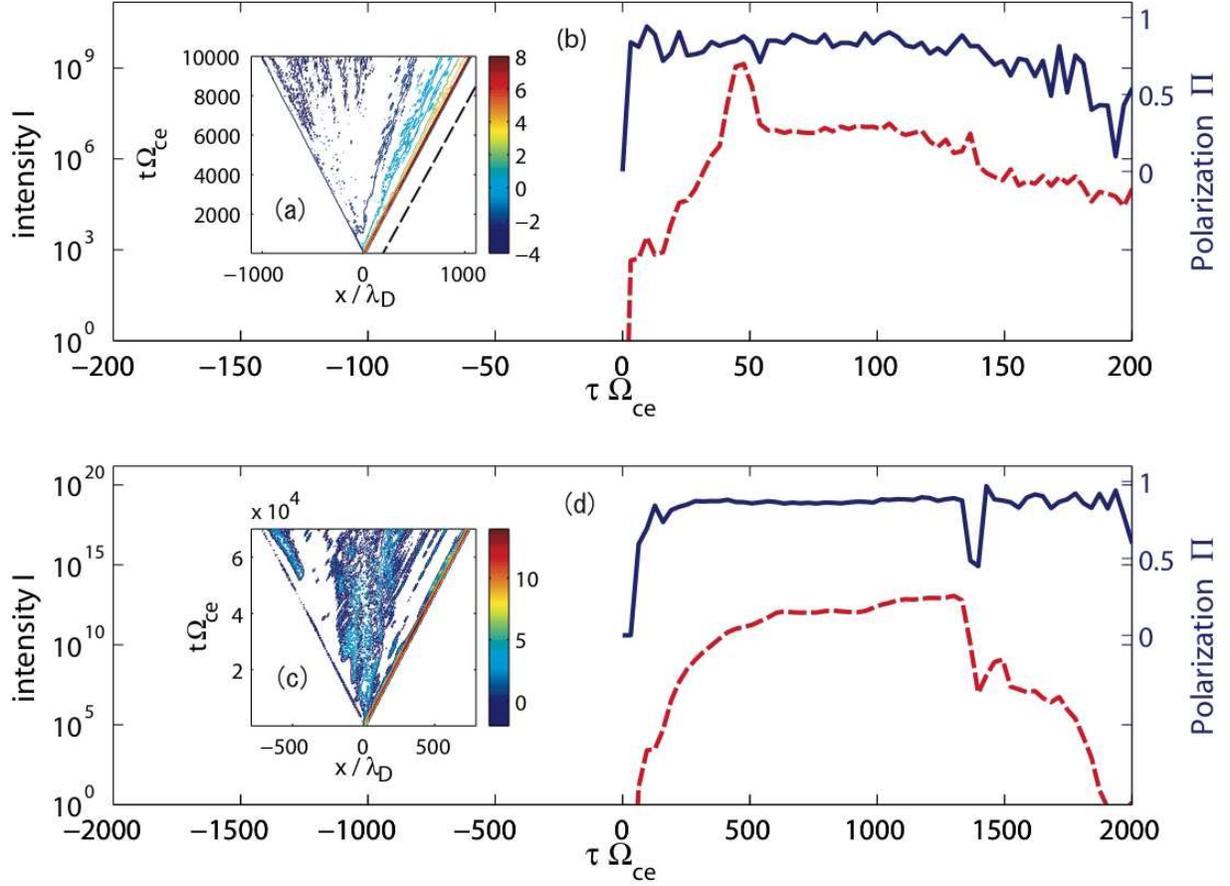}
\caption{\label{fig:int} Contour plots of instantaneous intensity $\log_{10} I(\theta=\phi=0)$ as functions of source frame coordinates $(x, t)$ for  for (a) Run C and (c) Run F. Intensity (dashed lines, left scales) and polarization 
(solid lines, right scales) as functions of detector time $\tau$ for (b) Run C and (d) Run F at $\theta=\phi=0$. 
Intensity scale is arbitrary. 
To obtain time dependence of intensity, instantaneous intensity from each particle
is summed up along the light cone $\tau=t-R/c=$const., shown as a black dashed line in panel (a). Note that actual detected pulses are much broader when we take into account angular effects (c.f. Fig. \ref{fig:itg})}
\end{figure}

The time dependence of detected intensity $log_{10}I$ from each particle along $\phi=\theta=0$ before the summation is shown as a contour plot in 
Fig.~\ref{fig:int}(a) and (c). Since the observer is in the positive $x$ direction, radiation comes from the particles near the front in the positive $x$ region.   
Since energetic particles
are bouncing back and forth within the ponderomotive potential well and slower particles are 
dropped off from the well, the radiation duration is broadened.  

Sample intensity $I$ is shown as solid lines in Fig.~\ref{fig:int}(b) and (d), indicating the detected
radiation peak width
is $\Delta t\sim 20 \Omega_{ce}^{-1}=2\lambda_D/c$ for Run C  and $\Delta t\sim800 \Omega_{ce}^{-1}=8\lambda_D/c$ for Run D.
This suggests that the radiation pulse width increases with $B$, but we emphasize that actual detected pulse are much broader due to angular effect. 
Polarization $\Pi$ is shown as dotted lines in Fig.~\ref{fig:int}(b) and (d),
confirming that the radiation is strongly linear-polarized as anticipated from the initial uniform magnetic field with the small depolarization coming from the initial random velocity distribution in the $y$ direction.

\begin{figure}
\includegraphics[width=\linewidth]{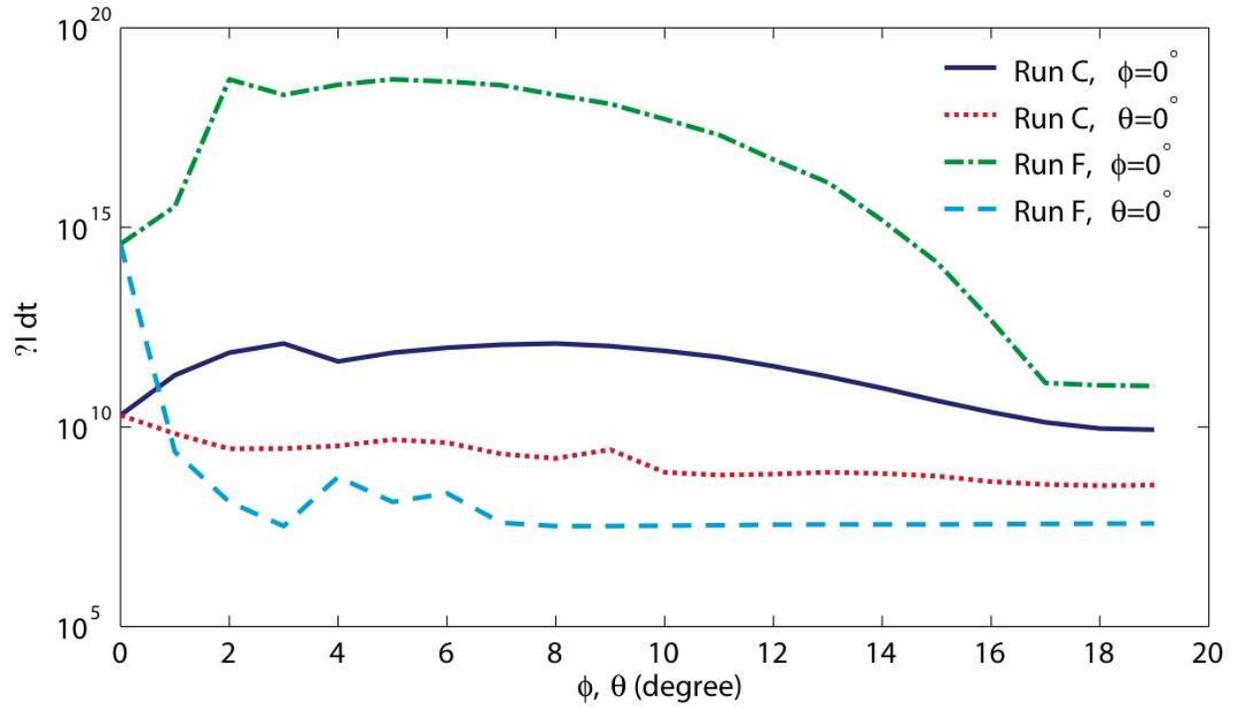}
\caption{\label{fig:itg} 
The total radiation angular fluence $\int I(\phi, \theta) d\tau$ for Runs C and F as a function of the 
angle ($\phi, \theta$). }
\end{figure}

In Fig.~\ref{fig:itg}, we show the total angular fluence $\int I(\phi, \theta) d\tau$
 as a function of the view angle. 
Angular fluence peaks around $\theta=3\sim 8^\circ$ in $\phi=0$ cases (solid and dash-dotted lines), 
corresponding to not the direction of the Poynting vector but
the direction of high energy particles in Fig.~\ref{fig:phs}. 
Angular fluence rapidly decreases with $\phi$ in $\theta=0$ case, indicating radiation is strongly collimated in the $x-z$ plane.

In summary, we observe the self-consistent radiation damping and radiation output from the acceleration  of electron-positron plasma via a relativistic PIC simulation. We find that the coupling between the field and particles becomes less efficient 
with larger radiation damping, implying the actual cooling time by the radiation increases. Radiation damping decelerates the energetic particles, but the ratio between $P_x$ and $P_z$ retains. The resulting radiation angular fluence peaks in the direction of the energetic particles. The radiation power monotonically decreases in the early time and becomes more or less constant later, and it is always smaller than the non-radiative case, indicating continuous cooling by the radiation damping. The radiation field is strongly linear-polarized both in
weak and strong magnetic field cases, which may be detectable by $\gamma$-ray burst observations as an indication of Poynting flux acceleration.
The simulations shown here are still too short to determine the cooling time when most of the EM energy is radiated away. 
Such questions remain to be answered by much longer runs.

This research is partially supported by NASA Grant No. NAG5-9223, NSF Grant No. AST0406882,  
and LLNL contract nos. B528326 and B541027. KN is also grateful to ILSA, LLNL for hospitality, and B. Remington and S. Wilks for useful discussions.

\newpage 

\newpage
\begin{table}
\caption{\label{table1}The parameters for each runs}
\begin{ruledtabular}
\begin{tabular}{cccc}
&$k$&$\omega_{pe}/\Omega_{ce}$& Duration $t\Omega_{ce}$\\
\hline
Run A & 0         & 0.1 & 10000\\
Run B & $10^{-4}$ & 0.1 & 10000\\
Run C & $10^{-3}$ & 0.1 & 10000\\
Run D & 0         & 0.01 & 70000\\
Run E & $10^{-4}$ & 0.01 & 70000\\
Run F & $10^{-3}$ & 0.01 & 70000\\
\end{tabular}
\end{ruledtabular}
\end{table}

\newpage

\newpage

\newpage

\newpage

\end{document}